\documentclass[3p,times,numbers,sort&compress]{elsarticle}
\usepackage{amssymb}
\usepackage{amsmath}
\journal{Chinese Journal of Physics}

\usepackage[colorlinks,citecolor=blue,linkcolor=blue,urlcolor=blue,hyperindex]{hyperref}
\usepackage{caption}
\usepackage{subcaption}
\usepackage{setspace}

\begin{document}

\begin{frontmatter}

\title{From Independent to Joint: Enhancing Quantum Phase and Correlation Factor Estimation by Squeezed Reservoir Engineering} %% Article title

\author[]{ \leftline{Cai-Hong Liao$^{1}$, Yan-Ling Li$^{1,2,*}$, Long Huang$^{1}$, Xing Xiao$^{3,**}$}}

\address{ \leftline {$^{1}$School of Information Engineering, Jiangxi University of Science and Technology, Ganzhou 341000, China}
 \leftline {$^{2}$Jiangxi Provincial Key Laboratory of  Multidimensional Intelligent Perception and Control, Jiangxi University of Science and Technology, Ganzhou 341000, China}
  \leftline {$^{3}$School of Physics and Electronic Information, Gannan Normal University, Ganzhou 341000, China}
}
\cortext[]{liyanling@jxust.edu.cn}
\cortext[]{xiaoxing@gnnu.edu.cn}

%% Abstract
\begin{abstract}
High-precision quantum parameter estimation is fundamental to the advancement of quantum metrology. Although reservoir engineering provides a powerful approach to improve estimation by tailoring system-environment interactions, the role of the squeezing phase and correlations arising from the sequential utilization of the same squeezed reservoir remains inadequately explored. In this work, we employ a correlated squeezed-thermal reservoir to enhance the precision of estimating the phase parameter $\phi$ and the correlation factor $\mu$, both individually and simultaneously. We show that the squeezing phase $\Phi$ is crucial for achieving quantum-enhanced precision, with optimal phase-matching conditions that depend strongly on $\mu$. Specifically, we derive the near-optimal phase-matching relations aimed at maximizing the quantum Fisher information (QFI) for both $\phi$ and $\mu$, as well as minimizing the total variance $\Delta_{\rm sim}$ in joint estimation. Furthermore, we show that the joint estimation variance is dominated by $F_{\phi}$, which motivates our search for the phase-matching conditions that minimize $\Delta_{\text{sim}}$. Through the ratio $R$ of variances, we demonstrate that joint estimation conserves quantum resources and maintains high precision when the squeezing phase is optimized for $F_{\phi}$, despite the inherent incompatibility of the parameters. These findings provide practical insights into reservoir engineering strategies for high-precision quantum sensing and information processing.
%While the enhancement of phase sensitivity is contingent upon satisfying these relations, the precision of estimating $\mu$ is universally improved by squeezing across all squeezing angles.
\end{abstract}

\begin{keyword}
Quantum Fisher information \sep Joint estimation \sep Correlated squeezed reservoir \sep Phase-Matching Conditions
\end{keyword}

\end{frontmatter}

\section{Introduction}
\label{intro}

Quantum parameter estimation seeks to achieve high-precision measurements of specified parameters by leveraging quantum resources \cite{Pez2018, Dan2018}. This approach has extensive applications across various research domains, including gravitational wave detection \cite{Cav1981, Gard2024}, atomic clocks \cite{Nichol2022, Zaheer2025}, and quantum imaging \cite{Tsang2009, Albare2020}. It has been established that, for closed quantum systems, the precision of parameter estimation can attain the standard quantum limit (SQL) when the probe state is separable \cite{Esc2011, Kol2013}. By employing quantum resources such as coherence, squeezing, or multipartite entanglement, the precision can be further enhanced, approaching the Heisenberg limit \cite{Pez2018, Str2017, Zwi2012}.

At the beginning of quantum parameter estimation theory, the majority of research efforts focused on single-parameter estimation, particularly concerning the phase parameter \cite{Braunstein1994, Toth2014}. However, in recent years, physicists have increasingly investigated the domain of multi-parameter estimation, both theoretically and experimentally \cite{Bagan2001, Crowle2014, Monras2011, Szczy2016, Demko2020}. The quantum Fisher information (QFI) and the quantum Fisher information matrix (QFIM) serve to characterize the ultimate precision limits of single-parameter and multi-parameter estimation, respectively, as delineated by the quantum Cram\'er-Rao bound \cite{Helstrom1969}.

However, it is crucial to recognize that any realistic physical system inevitably interacts with its environment, which leads to a degradation in the precision of parameter estimation \cite{Degen2017, Nielsen2000}. The challenge of mitigating the impact of environmental noise on quantum systems has emerged as a significant topic in the field of quantum-enhanced technologies.
In recent decades, it has been widely recognized that the environment invariably exerts a detrimental influence on quantum parameter estimation. Consequently, numerous techniques have been proposed to alleviate the effects of environmental noise and enhance measurement precision. Such methods include dynamical decoupling \cite{Viola1999, Lange2010}, quantum error correction \cite{Chiaverini2004, Kessler2014}, weak measurement \cite{Sun2010, Kim2011, Li2023}, and environment-assisted measurement \cite{Goldstein2011, Zhao2013}. These approaches illustrate both the advancements and ongoing challenges in the realm of quantum parameter estimation. Nonetheless, they also possess limitations that constrain their effectiveness. For instance, dynamical decoupling may prove less effective against high-frequency noise or noise characterized by a broad spectrum. Additionally, both weak and environment-assisted measurements operate as probabilistic schemes, wherein the corresponding probabilities may diminish in regions where they could otherwise provide significant improvements in the precision of parameter estimation.

On the contrary, reservoir engineering refers to the manipulation and control of quantum systems by tailoring their interactions with an environment or reservoir, representing a powerful tool in quantum information science \cite{Muessel2014, Adesso2016, Jeong2019, Xu2019, Gessner2020, Wang2022, Lan2023}. This work explores how parameter estimation precision can be enhanced by engineering the squeezing properties of the reservoir, as well as by exploiting correlations resulting from consecutive use of the same squeezed reservoir.
Unlike previous studies that primarily focused on the relationship between squeezing strength and quantum entanglement or spin squeezing \cite{Woolley2014, Kienzler2015, Yang2015, Didier2018, Hou2018, Hou2019, Bai2021, Groszkowski2022, LiY2023}, our work highlights the importance of the squeezing phase and the correlation effects in quantum parameter estimation. We demonstrate that the conditions for phase-matching play a pivotal role in quantum parameter estimation, providing the potential to surpass the standard quantum limit \cite{Cav1981, Liu2013}.

We investigate independent and joint estimation of the phase parameter $\phi$ and the correlation factor $\mu$. It is important to clarify the physical motivation for estimating the phase parameter $\phi$ and the correlation factor $\mu$. On the one hand, the phase parameter $\phi$ represents the most common unknown quantity of interest in metrological tasks. Our framework is general and unifies two primary physical scenarios: (i) Quantum sensing where a known probe state (e.g., with $\phi=0$) acquires the unknown phase $\phi$ through interaction with an external process before it enters a noisy reservoir, or (ii) Quantum state characterization where the quantum source itself produces an entangled state with an unknown, intrinsic phase $\phi$ that we seek to characterize. Our mathematical framework encompasses both scenarios, which are fundamental tasks in quantum metrology. On the other hand, the scarcity of literature on estimating the correlation factor motivates our focus on achieving high-precision metrology for this quantity. Moreover, joint estimation of $\phi$ and $\mu$ presents distinct advantages over independent schemes, including reduced quantum resource consumption and progress in multiparameter quantum metrology.

In this paper, we demonstrate that the enhancement from squeezing is critically dependent on the phase-matching condition between the squeezing phase $\Phi$ and the estimated phase $\phi$. We find that this optimal condition itself is not fixed, but rather depends directly on the correlation factor $\mu$ of the channel. We derive the specific conditions for both weak ($\mu \to 0$) and strong ($\mu \to 1$) correlation regimes, revealing the intricate interplay between these two engineered environmental properties. The enhancement of the precision of $\phi$ through squeezing is contingent upon the fulfillment of the phase-matching conditions. Failure to meet these conditions may lead to a degradation in estimation precision. In contrast, it is worth emphasizing that squeezing can enhance the precision of $\mu$ across all squeezing phases, though optimal enhancement still requires phase-matching.
Furthermore, we investigate the effect of the phase-matching conditions on the lower bound of the total variance $\Delta_{\rm sim}$ in the context of joint estimation of the parameters $\phi$ and $\mu$. The results reveal that the lower bound of $\Delta_{\rm sim}$ is primarily determined by the QFI $F_{\phi}$. Consequently, under the phase-matching conditions that maximize $F_{\phi}$, $\Delta_{\rm sim}$ is significantly minimized, facilitating higher joint estimation precision. Finally, we compare the performance of joint and independent estimation protocols via the ratio $R$, a metric that characterizes the merits of joint estimation. We analyze how phase-matching affects this metric. This study contributes to a deeper understanding of the operational requirements for employing squeezing and correlation effects in advancing quantum parameter estimation, thereby offering valuable insights for experimental implementations.

Our paper is organized as follows. In Sec. \ref{model}, we show the theoretical model and derive the analytical expression of the reduced density matrix of the two-qubit entangled state. In Sec. \ref{phase} and Sec. \ref{mu}, we show that the precisions of $\phi$ and $\mu$ can be enhanced by engineering the
squeezing properties of the reservoir, as well as by exploiting input correlations resulting from repeated use of
the same squeezed reservoir. In particular, we derive near-optimal phase-matching
relations aimed at maximizing $F_{\phi}$ and $F_{\mu}$. In Sec. \ref{joint}, we discuss the joint estimation of $\phi$ and $\mu$, and
we demonstrate that joint estimation conserves quantum resources
and maintains high precision. In Sec. \ref{conclusions}, a summary is given.

\section{Theoretical Model}
\label{model}
\begin{figure}[htbp]
\centering
\includegraphics[width = 16 cm]{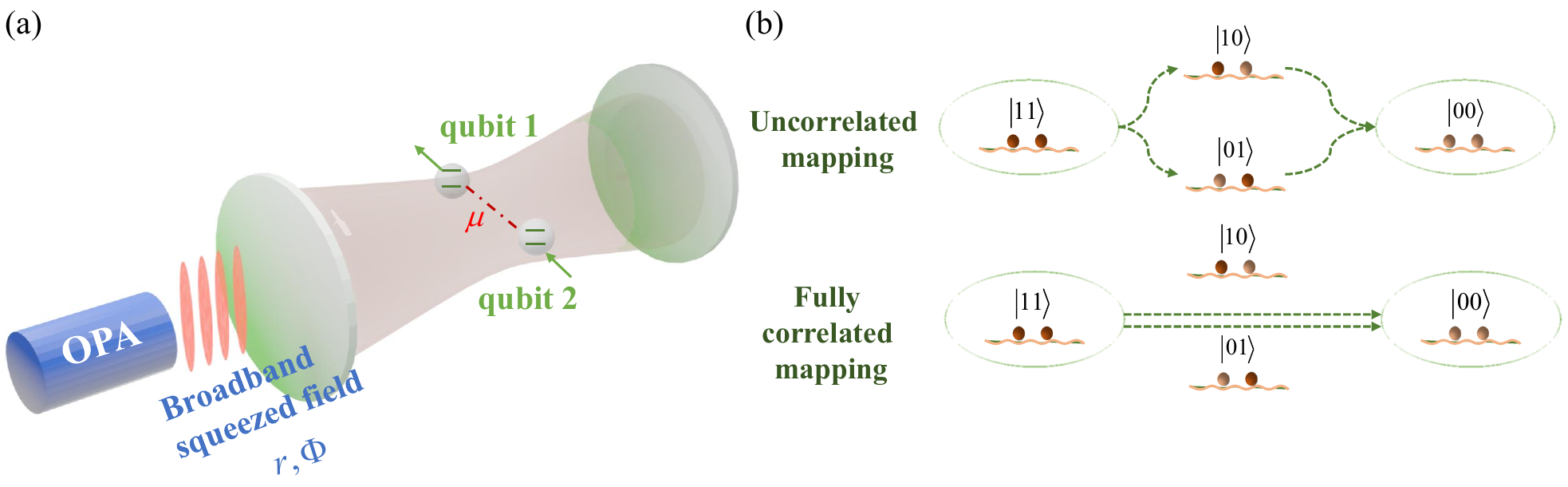}
\caption{(color online) (a) The theoretical model of two initially entangled qubits passing through the squeezed thermal reservoir successively at time intervals $\Delta{t}$. If the relaxation time ${t}_{rel}$ of the cavity is not significantly shorter than the time interval $\Delta{t}$, the influence of the reservoir on the two qubits will be correlated with probability $\mu$. (b) Mapping diagrams for a two-qubit system under uncorrelated and fully correlated conditions.}
\label{fig1}
\end{figure}
\label{theor}

Our physical model consists of two two-level systems (TLS), labeled as qubits 1 and 2, interacting with a common bosonic reservoir. The general Hamiltonian in the rotating-wave approximation is ($\hbar= 1$): $H_{\text{total}} = H_S + H_R + H_I$ with
\begin{align}
H_S &= \sum_{i=1,2} \frac{\omega_0}{2} \sigma_z^i,\\
H_R &= \sum_k \omega_k a_k^\dagger a_k,\\
H_I &= \sum_{i=1,2} \sum_k g_{k} (\sigma_+^i a_k + \sigma_-^i a_k^\dagger),
\end{align}
where $\sigma_z^i$ is the Pauli operator of the $i$th qubit with transition frequency $\omega_{0}$, $a_{k}$ is the annihilation operator of the $k$th field mode with frequency $\omega_{k}$, and $g_{k}$ is the coupling strength. For simplicity, we have assumed the two qubits are identical.

The specific dynamics of the two-qubit system are determined by the state of the reservoir and the nature of the interaction.
We consider two qubits passing through the cavity in succession and perform reservoir engineering by assuming the reservoir is not a simple thermal reservoir but a squeezed thermal reservoir. As shown in Fig. \ref{fig1}(a), the squeezed thermal reservoir is realized by injecting a broadband squeezed field into the optical cavity. The function of the optical parametric amplifier (OPA) is to generate the broadband squeezed field. Consequently, this process effectively converts the environment of the cavity into a squeezed thermal reservoir \cite{Kimble1998}. Mathematically, the density matrix of this engineered reservoir, $\rho_R$, is modeled by applying the squeezing operators, $S_{k}(r,\Phi)=\exp[r(e^{-i\Phi} a_{k}^{2}-e^{i\Phi} a_{k}^{\dagger 2})/2]$, to a thermal state $\rho_{\rm th}$ (with mean photon number $n$): $\rho_R = \Pi_{k}S_{k}(r,\Phi) \rho_{\rm th} S_{k}^\dagger(r,\Phi)$, where $r$ is the squeezing strength, and $\Phi$ is the squeezing phase.

However, the majority of previous investigations have focused on the memoryless utilization of the cavity, wherein the noise sources are assumed to be independent. While this assumption may be reasonable in certain specific physical scenarios, it may not hold in many practical situations. Specifically, a high-Q cavity can retain memory effects, influencing the second qubit after interacting with the first. These correlation effects become crucial when the relaxation time of the cavity ($t_{rel}$) is not significantly shorter than the time interval ($\Delta t$) between the arrival of the two consecutive qubits. Therefore, a physical model capable of describing this ``partial memory'' is required.

To construct this model, we employ a well-established phenomenological framework. The central idea is to describe the complex ``partial memory'' dynamics as a probabilistic interpolation between two distinct limiting cases, as depicted in Fig.~\ref{fig1}(b):

\textbf{Limit 1: Uncorrelated ($\mu=0$).} This corresponds to the memoryless limit ($t_{rel} \ll \Delta t$), where the two qubits interact independently with the squeezed reservoir.

\textbf{Limit 2: Fully Correlated ($\mu=1$).} This corresponds to the perfect memory limit ($t_{rel} \gg \Delta t$), where the two qubits interact collectively with the same squeezed reservoir, necessitating a description via joint operators.

We thus introduce the correlation factor $\mu$ ($0 \leq \mu \leq 1$) to quantify the strength of this memory effect. The total evolution of the system is then modeled as a convex mixture of these two mappings: the system undergoes the uncorrelated evolution with probability $(1-\mu)$ and the fully correlated evolution with probability $\mu$. Consequently, the final output state $\rho(t)$ can be expressed as:
\begin{equation}
  \rho (t)=(1-\mu ){{\rho }^{(u)}}(t)+\mu {{\rho }^{(c)}}(t),
  \label{eq4}
\end{equation}
where $\rho^{(u)}(t)$ describes the uncorrelated part, while $\rho^{(c)}(t)$ describes the fully correlated one.

For the uncorrelated part, qubits 1 and 2 are independently affected by the squeezed thermal reservoir. Under the Born-Markov approximation, the time evolution of the reduced density matrix is governed by the following master equation \cite{Srikanth2008}
\begin{eqnarray}
\label{eq5}
\dot{\rho}^{(u)}(t)&=&\sum_{i=1}^2\Big( -\frac{\gamma_i N}{2}\left[\{\sigma_-^i\sigma_+^i,\rho_u(t)\}-2\sigma_+^i\rho^{(u)}(t)\sigma_-^i\right]
-\frac{\gamma_i(N+1)}{2}\left[\{\sigma_+^i\sigma_-^i,\rho^{(u)}(t)\}-2\sigma_-^i\rho^{(u)}(t)\sigma_+^i\right] \nonumber\\
&&  -\gamma_i\chi e^{i\Phi}\sigma_+^i\rho^{(u)}(t)\sigma_+^i -\gamma_i\chi e^{-i\Phi}\sigma_-^i\rho^{(u)}(t)\sigma_-^i)\Big),
\end{eqnarray}
where $\{\cdot, \cdot\}$ denotes the anti-commutator, ${{\sigma }_{\pm}^{1}}={{\sigma }_{\pm}}\otimes I$, ${{\sigma }_{\pm}^{2}}=I \otimes {{\sigma }_{\pm}}$, and $\sigma_{+}=|1\rangle\langle0|$, $\sigma_{-}=|0\rangle\langle1|$. Here, $\gamma_{i}$ is the spontaneous emission rate of the TLS, and without loss of generality, we suppose $\gamma_{1}=\gamma_{2}=\gamma_{0}$. The effective photon number is given by $N=n\cosh2r+\sinh^{2}r$, where the mean photon number $n= 1/\left[\exp({\frac{\hbar \omega_{k}}{k_{\small{B}} T}})-1)\right]$ and $T$ denotes the temperature of the squeezed-thermal reservoir. The squeezing parameter is defined as $\chi=-\frac{1}{2}(2n+1)\sinh2r$.

However, if the time interval between two qubits passing through the cavity is extremely short, then the consecutive uses of the reservoir are almost perfectly correlated. In this case, the dynamical map cannot be expressed as the tensor product of the maps of the individual qubits, but is described by the correlated Lindblad equation \cite{Yeo2003}
\begin{eqnarray}
  \label{eq6}
\dot{\rho}^{(c)}(t) &=&-\frac{{{\gamma }_{0}}N}{2}\left[\{\sigma _{-}^{\otimes 2}\sigma _{+}^{\otimes 2},\rho^{(c)}(t)\}-2\sigma _{+}^{\otimes 2}\rho^{(c)}(t)\sigma _{-}^{\otimes 2}\right]
   -\frac{{{\gamma }_{0}}(N+1)}{2}\left[\{\sigma _{+}^{\otimes 2}\sigma _{-}^{\otimes 2},\rho^{(c)}(t)\}-2\sigma _{-}^{\otimes 2}\rho^{(c)}(t)\sigma _{+}^{\otimes 2}\right] \nonumber \\
 && -{{\gamma }_{0}}\chi {{e}^{i\Phi }}\sigma _{+}^{\otimes 2}\rho^{(c)}(t)\sigma _{+}^{\otimes 2}-{{\gamma }_{0}}\chi {{e}^{-i\Phi }}\sigma _{-}^{\otimes 2}\rho^{(c)}(t)\sigma _{-}^{\otimes 2},
\end{eqnarray}
where $\sigma _{\pm}^{\otimes 2}=\sigma _{\pm} \otimes \sigma _{\pm}$.

We assume the initially entangled state is given by
\begin{equation}
  |\psi\rangle_{12}={\alpha|00\rangle}_{12}+ e^{i\phi}{\beta|11\rangle}_{12},
  \label{eq7}
\end{equation}
with $\alpha^{2}+\beta^{2}=1$. The preparation of such entangled states is a standard resource for quantum information protocols and has been experimentally demonstrated with high fidelity in various physical platforms, such as superconducting qubits \cite{Blais2021, Barends2014} or trapped ions \cite{Bruzewicz2019, Ballance2016}. As we elucidated in the introduction, the unknown phase parameter $\phi$ may arise from interactions with an external process or from an intrinsic phase acquired during the state preparation.

The specific form of $\rho(t)$ in Eq. (\ref{eq4}) can be obtained by solving the above master equation in Eqs. (\ref{eq5}) and (\ref{eq6}), which yields to be
\begin{equation}
  \rho (t)=\left( \begin{matrix}
   {{\rho }_{11}}(t) & 0 & 0 & {{\rho }_{14}}(t)  \\
   0 & {{\rho }_{22}}(t) & {{\rho }_{23}}(t) & 0  \\
   0 & {{\rho }_{32}}(t) & {{\rho }_{33}}(t) & 0  \\
   {{\rho }_{41}}(t) & 0 & 0 & {{\rho }_{44}}(t)  \\
\end{matrix} \right).
\label{eq8}
\end{equation}

To clarify the physical contributions from the uncorrelated and correlated evolutions as defined in Eq. (\ref{eq4}), we restructure the dense expressions for the matrix elements as follows:
\begin{subequations}
\begin{align}\label{eq9a}
    \rho_{11}(t) &= \overline{\mu} \rho_{11}^{(u)} + \mu \rho_{11}^{(c)},  \nonumber\\
    \rho_{44}(t) &= \overline{\mu} \rho_{44}^{(u)} + \mu \rho_{44}^{(c)},  \nonumber \\
    \rho_{22}(t) &= \rho_{33}(t) = \overline{\mu} \rho_{22}^{(u)},  \\
    \rho_{14}(t) &= \rho_{41}^{*}(t) = \overline{\mu} \rho_{14}^{(u)} + \mu \rho_{14}^{(c)},  \nonumber\\
    \rho_{23}(t) &= \rho_{32}^{*}(t) = \overline{\mu} \rho_{23}^{(u)}, \nonumber
\end{align}
where the uncorrelated contributions ($\rho_{ij}^{(u)}$) are:
\begin{align}\label{eq9b}
    \rho_{11}^{(u)} &= \frac{1}{(2N+1)^{2}} \left[ (1+N)^{2} - 2(1+N)Qd^{2} + Vd^{4} \right],  \nonumber\\
    \rho_{44}^{(u)} &= \frac{1}{(2N+1)^{2}} \left[ N^{2} + 2NQd^{2} + Vd^{4} \right],  \nonumber\\
    \rho_{22}^{(u)} &= \frac{1}{(2N+1)^{2}} \left[ N(1+N) + Qd^{2} - Vd^{4} \right],  \\
    \rho_{14}^{(u)} &= \alpha\beta d^{2} \left[ e^{-i\phi}\cosh^{2}(\chi\gamma_{0}t) + e^{i(\phi-2\Phi)}\sinh^{2}(\chi\gamma_{0}t) \right], \nonumber\\
    \rho_{23}^{(u)} &= -\alpha\beta d^{2}\cos(\phi-\Phi)\sinh(2\chi\gamma_{0}t), \nonumber
\end{align}
and the fully correlated contributions ($\rho_{ij}^{(c)}$) are:
\begin{align}\label{eq9c}
    \rho_{11}^{(c)} &= \frac{1}{2N+1} (N+1-Qd^{2}),  \nonumber\\
    \rho_{44}^{(c)} &= \frac{1}{2N+1} (N+Qd^{2}),  \\
    \rho_{14}^{(c)} &= \alpha\beta d \left[ e^{-i\phi}\cosh(\chi\gamma_{0}t) - e^{i(\phi-\Phi)}\sinh(\chi\gamma_{0}t) \right], \nonumber
\end{align}
\end{subequations}
with $Q=(2N+1){{\beta }^{2}}-N$, $V=(2N+1){{\beta }^{2}}+{{N}^{2}}$, $d=e^{-(2N+1)\gamma_{0}t/{2}}$, and $\bar{\mu }=1-\mu$. Note that Eqs. (\ref{eq9a}-\ref{eq9c}) is the analytical time-dependent solution for the density matrix $\rho(t)$, which describes the transient dynamics of the system.

This analytical expression enables us to estimate any parameters $\xi ={{({{\xi }_{1}},\cdots ,{{\xi }_{a}},\cdots ,{{\xi }_{b}},\cdots )}^{T}}$ included in Eq. (\ref{eq8}). The performance of the estimator can be quantified using the covariance matrix ${\rm Cov}(\xi)$, which captures not only the variances of individual parameters, but also the covariances between them. For the unbiased estimator $\hat{\xi }={{({{\hat{\xi }}_{1}},\cdots ,{{\hat{\xi }}_{a}},\cdots ,{{\hat{\xi }}_{b}},\cdots )}^{T}}$, the quantum Crm\'er-Rao bound establishes a lower bound on the covariance matrix, which is determined by the QFIM \cite{Braunstein1994}
\begin{equation}\label{eq10}
  \operatorname{cov}(\hat{\xi })\ge {{F}^{-1}},
\end{equation}
where $\operatorname{cov}(\hat{\xi })$ represents the covariance matrix, and ${{F}}$ is the QFIM. In scenarios where the estimation is restricted to two parameters, $\xi_a$ and $\xi_b$, the explicit forms of the covariance matrix and the QFIM are given by:
\begin{equation}\label{eq11}
  \operatorname{cov}(\hat{\xi }) =\left( \begin{matrix}
   \operatorname{var}(\hat{\xi }_{a}) & \operatorname{cov}(\hat{\xi }_{a},\hat{\xi }_{b})  \\
   \operatorname{cov}(\hat{\xi }_{b},\hat{\xi }_{a}) & \operatorname{var}(\hat{\xi }_{b})   \\
\end{matrix} \right), \quad
   F =\left( \begin{matrix}
   F_{\xi_{a}\xi_{a}} & F_{\xi_{a}\xi_{b}}  \\
   F_{\xi_{b}\xi_{a}} & F_{\xi_{b}\xi_{b}}   \\
\end{matrix} \right),
\end{equation}
where $\operatorname{var}({{\hat{\xi }}_{a}})=E\left[ {{({{{\hat{\xi }}}_{a}}-{{\xi }_{a}})}^{2}} \right]$, and $\operatorname{cov}({{\hat{\xi }}_{a}},{{\hat{\xi }}_{b}})=E\left[ ({{{\hat{\xi }}}_{a}}-{{\xi }_{a}})({{{\hat{\xi }}}_{b}}-{{\xi }_{b}}) \right]$. It is worth noting that the diagonal elements of the QFIM correspond to the QFI of the individual parameters (i.e., $F_{\xi_{a}}=F_{\xi_{a}\xi_{a}}$), whereas the off-diagonal terms characterize the correlations between them. The specific analytical expression for the cross-term $F_{\xi_{a}\xi_{b}}$ is derived in Ref.~\cite{Liu2020}
\begin{equation}
 F_{\xi_{a}\xi_{b}}=\sum\limits_{i,j=0}^{3}{\frac{2\operatorname{Re}(\left\langle  {{\lambda }_{i}} \right|{{\partial }_{{{{\xi }}_{a}}}}\rho \left| {{\lambda }_{j}} \right\rangle \left\langle  {{\lambda }_{j}} \right|{{\partial }_{{{{\xi }}_{b}}}}\rho \left| {{\lambda }_{i}} \right\rangle )}{{{\lambda }_{i}}+{{\lambda }_{j}}}},
  \label{eq12}
\end{equation}
where $\lambda_i$ and $|\lambda_i\rangle$ denote the eigenvalues and eigenstates of the two-qubit reduced density matrix, respectively.

\section{Enhancing the Independent Estimation of Quantum Phase}
\label{phase}
In this section, we discuss how to enhance the precision of the phase parameter $\phi$ via the engineering of
the squeezed reservoir and correlation effects. According to Eq. (\ref{eq12}), the QFI $F_\phi$ can be obtained analytically.
However, we must first note that the full expression for $F_{\phi}$, given in Eq.~(\ref{eq13}) below, is explicitly complex. This complexity arises because $F_{\phi}$ is a highly non-linear function of the density matrix elements, critically depending on the system's eigenvalues $\lambda_i$ and eigenstates $|\lambda_i\rangle$. Fortunately, as we will show, we can extract the essential physics by adopting this exact decompositional logic on the most critical term containing the phase information: the coherence $|\rho_{41}|^2$. The QFI of the phase parameter $\phi$ is
\begin{equation}
\label{eq13}
  \begin{aligned}
    F_{\phi}=&K+\frac{{{\left| {{\rho }_{\text{41}}} \right|}^{\text{2}}}\cdot 4\left[ \left(\rho_{11}\rho_{44}-|\rho_{41}|^{2}\right){{\left| {{D}_{\phi }} \right|}^{\text{2}}}+{{\left| {{\rho }_{\text{41}}} \right|}^{\text{2}}}{[\operatorname{Re}(D_{\phi})]^{2}} \right]}{(\rho_{11}+\rho_{44})\left(\rho_{11}\rho_{44}-{{\left| {{\rho }_{\text{41}}} \right|}^{\text{2}}}\right)},
  \end{aligned}
\end{equation}
where
\begin{align}
 \label{eq14}
  K &= \frac{2\rho_{22}G_{0}\bigl[1-\cos(2\Phi-2\phi)\bigr]}
            {2\rho_{22}^{2}-G_{0}\bigl[1+\cos(2\Phi-2\phi)\bigr]}, \quad
             {{G}_{0}}={{\bar{\mu }}^{2}}{{\alpha }^{2}}{{\beta }^{2}}{{d}^{4}}{{\sinh }^{2}}(2\chi {{\gamma }_{0}}t),\quad
  %S &= (\rho_{11}-\rho_{44})^2 + 4|\rho_{41}|^{2}, \quad
  D_{\phi} = \frac{\partial_{\phi}\rho_{41}}{\rho_{41}}.
\end{align}
The notation ${{\partial }_{\phi }}$ serves as an abbreviation for $\frac{{\partial }}{{\partial }{\phi }}$. In the subsequent discussion, we adopt $T \rightarrow \widetilde{T} \equiv \frac{k_{B}T}{\hbar\omega_{k}}$. For the sake of simplicity, we will continue to denote $\widetilde{T}$ as $T$.

\begin{figure*}[htbp]
	\centering
\includegraphics[width = 16 cm]{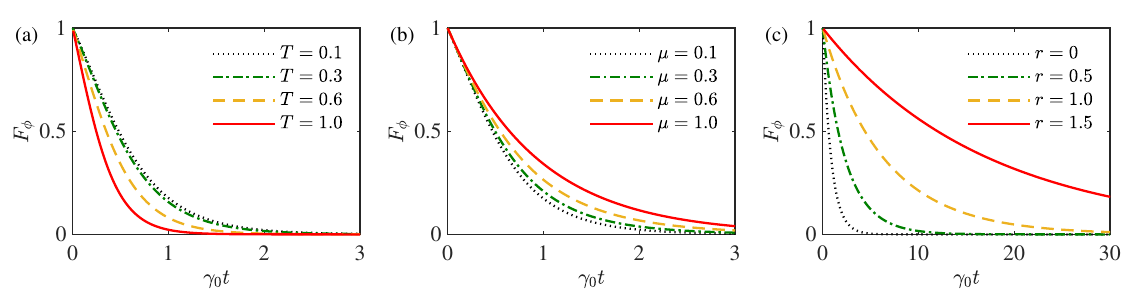}
    \caption{(color online) $F_{\phi}$ as a function of $\gamma_{0}t$ with (a) $\mu=0$, $r={0}$, (b) $T={0.3}$, $r=0$, and (c) $T=0.3$, $\mu={0.9}$. The other parameters are $\Phi=0$, $\alpha=\frac{\sqrt{2}}{2}$, and $\phi=\pi/2$.}
	\label{fig2}
\end{figure*}

The numerical results for $F_{\phi}$ are presented in Fig. {\ref{fig2}}. In the absence of reservoir squeezing and correlation effects, the interaction with a thermal reservoir leads to a rapid decay of $F_\phi$ to zero, with higher temperatures accelerating this decay, as shown in Fig. {\ref{fig2}}(a). In contrast, the introduction of reservoir squeezing and correlation effects can significantly mitigate the decay of $F_\phi$. As depicted in Fig. {\ref{fig2}}(b), we illustrate the decay of $F_\phi$ with respect to $\gamma_{0}t$ for various correlation factors $\mu$. It is evident that stronger correlations more effectively preserve $F_\phi$, indicating a positive role of correlated noise in the squeezed reservoir. Similarly, the decay of $F_\phi$ with $\gamma_{0}t$ for different squeezing strengths $r$ is analyzed in Fig. {\ref{fig2}}(c).
The results demonstrate that increased squeezing strength significantly enhances $F_\phi$ at all timescales, with greater squeezing strength yielding a more pronounced enhancement of $F_\phi$.
The efficiency of preservation scales with $r$, thereby confirming quantum noise redistribution as the underlying mechanism. This occurs because squeezing induces anisotropic transverse dissipation: noise in one quadrature is suppressed while that in the conjugate quadrature is amplified, which subsequently enhances phase sensitivity.

\begin{figure}[htbp]
	\centering
\includegraphics[width = 16 cm]{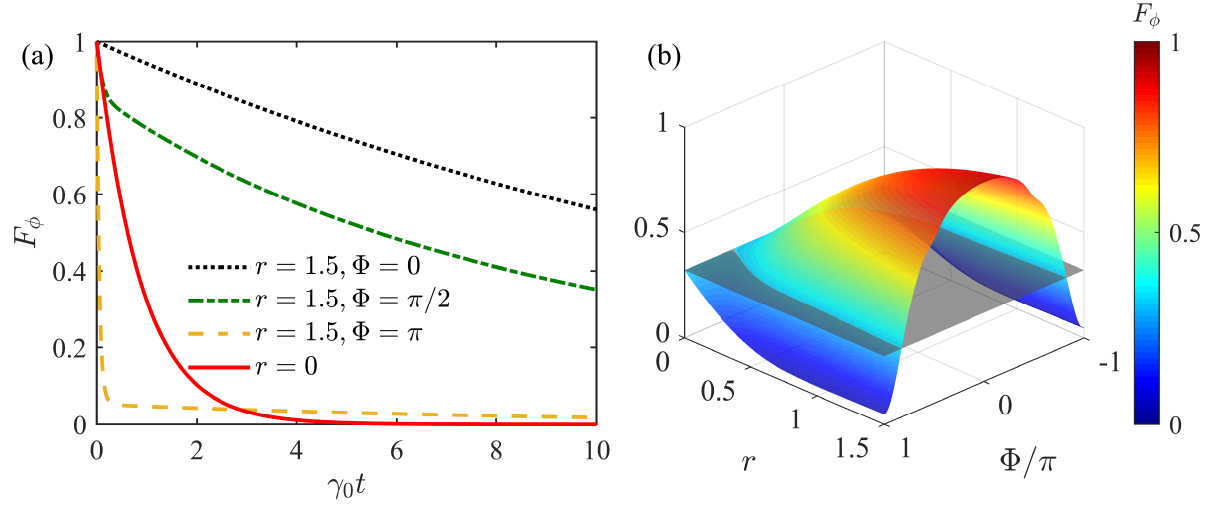}
 \caption{(color online) The behaviors of $F_{\phi}$ in different cases. (a) $F_{\phi}$ as a function of $\gamma_{0}t$ for different squeezed phases $\Phi$. (b) The coloured surface represents $F_{\phi}$ as a function of $\Phi$ and $r$ at $\gamma_{0}t=1$, while the grey plane serves as the reference plane for the results in a correlated thermal reservoir (without squeezing). The other parameters are $T=0.3$, $\mu={0.9}$, $\alpha=\frac{\sqrt{2}}{2}$, and $\phi=\pi/2$.}
	\label{fig3}
\end{figure}

A more detailed analysis reveals that the enhancement of $F_\phi$ through squeezing depends critically both on the squeezing strength $r$ and the squeezing phase $\Phi$. As illustrated in Fig. {\ref{fig3}}(a), an appropriate choice of squeezing phase leads to a greatly enhanced $F_\phi$, whereas a mismatched squeezing phase not only diminishes the beneficial effect but can even render $F_\phi$ lower than that of the unsqueezed case, despite large squeezing strength.

To further elucidate the role of the squeezing phase, Fig. \ref{fig3}(b) presents $F_\phi$ as a function of $r$ and $\Phi$. For comparison, the result for a correlated thermal reservoir (without squeezing) is included as the reference plane. It is clearly observed that, with a properly chosen $\Phi$, the values of $F_\phi$ in the correlated squeezed thermal reservoir lie above this reference plane, confirming the enhancement induced by squeezing. In contrast, when $\Phi$ is chosen inappropriately, the corresponding values fall below the reference plane, indicating that squeezing not only fails to improve precision but can even accelerate the decay of $F_\phi$. These results strongly suggest the existence of an optimal phase-matching condition that maximizes the beneficial effect of squeezing on $F_\phi$.

In order to specify the phase-matching conditions associated with $F_\phi$ in the context of a correlated squeezed thermal reservoir, a more detailed analysis of $F_\phi$ is required. According to Eq. (\ref{eq13}), we find that the squeezing phase $\Phi$ and the estimated phase $\phi$ are encoded within the parameters $K$ and ${{\left| {{\rho }_{41}} \right|}^{2}}$. The expression for $K$ has already been provided in Eq. (\ref{eq14}). We focus our analysis on the most physically significant term that encodes the phase $\phi$: the coherence $|\rho_{41}|^2$. This key term can be perfectly decomposed into the three physical contributions:
\begin{eqnarray}\label{eq15}
  &&{{\left| {{\rho }_{41}} \right|}^{2}} = {{\alpha }^{2}}{{\beta }^{2}}\cdot ({{\bar{\mu }}^{2}}{{G}_{1}}+{{\mu }^{2}}{{G}_{2}}+2\bar{\mu }\mu {{G}_{3}}),
\end{eqnarray}
where
\begin{eqnarray}\label{eq16}
 && {{G}_{1}} = {{x}^{4}}+{{y}^{4}}+2{{x}^{2}}{{y}^{2}}\cos (2\Phi -2\phi ),\nonumber\\
  &&{{G}_{2}} = {{x}^{2}}+{{y}^{2}}-2xy\cos (\Phi -2\phi ),\\
  &&{{G}_{3}} = {{x}^{3}}-{{x}^{2}}y\cos (\Phi -2\phi )\text{+}x{{y}^{2}}\cos (2\Phi -2\phi )-{{y}^{3}}\cos \Phi,\nonumber
\end{eqnarray}
with $x = d\cosh (\chi {{\gamma }_{0}}t)$, $y = d\sinh (\chi {{\gamma }_{0}}t)$.

Despite the complexity of these expressions, we can categorize them into three distinct classes:

(i) Uncorrelated terms governed by $\bar{\mu}^2$, such as $K$ and $G_1$;

(ii) Correlated terms dominated by ${\mu}^2$, including $G_2$;

(iii) Mixed terms governed by $\bar{\mu}\mu$, including $G_3$.

This classification provides a simple framework for analyzing the contributions of each term to $F_\phi$. Although there is no general phase-matching condition due to the complexity of the expression for the QFI, we can find two specific phase relations in the limiting cases of $\mu$.

In the regime of small $\mu$ ($\mu\rightarrow0$), the QFI $F_\phi$ is mainly determined by the uncorrelated terms. The near-optimal phase condition that allows $F_\phi$ to take a maximum value is
\begin{equation}
  \cos(2\Phi-2\phi)=-1.
\label{eq17}
\end{equation}

Similarly, in the regime of large $\mu$ ($\mu\rightarrow1$), the QFI $F_\phi$ is mainly determined by the correlated terms. The near-optimal phase condition that allows $F_\phi$ to take a maximum value is
\begin{equation}
 \cos(\Phi-2\phi)=-1.
 \label{eq18}
\end{equation}
Notice that Eqs. (\ref{eq17}) and (\ref{eq18}) are rigorous in the uncorrelated ($\mu=0$) and fully correlated ($\mu=1$) cases, respectively.

%
%Based on our comprehensive analysis, for the expression $K$, the condition that allows $K$ to take a maximum value can be derived by solving equation $\frac{\partial {{F}_{\phi }}}{\partial (2\Phi -2\phi )}=0$ and $\frac{{{\partial }^{2}}{{F}_{\phi }}}{\partial {{(2\Phi -2\phi )}^{2}}}<0$. Then we can obtain the condition is
%\begin{equation}
%  2\Phi-2\phi=\pm\pi.
%\end{equation}

\begin{figure}[htbp]
	\centering
\includegraphics[width = 16 cm]{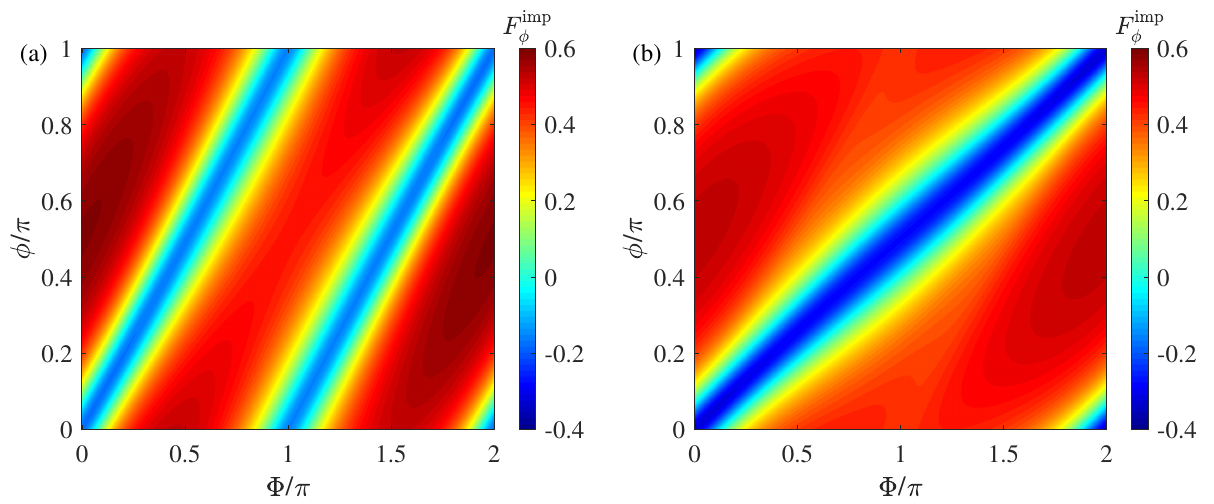}
    \caption{(color online) $F_{\phi}^{\rm imp}$ as a function of $\Phi$ and $\phi$ with (a) $\mu=0.1$ and (b) $\mu=0.9$. The other parameters are $\gamma_{0}t=1$, $T=0.3$, $r=1$, and $\alpha=\frac{\sqrt{2}}{2}$.}
	\label{fig4}
\end{figure}

To clarify the significant influences of the phase-matching conditions on the squeezing enhancement of parameter estimation, we introduce the QFI improvement $F_\phi^{\rm imp}$, which is defined as
 \begin{equation}
  F_\phi^{\rm imp}=F_\phi-F_\phi(r=0),
  \label{eq19}
\end{equation}
where $F_\phi(r=0)$ denotes the QFI without accounting for squeezing effects. The numerical results depicted in Figs. \ref{fig4}(a) and \ref{fig4}(b) illustrate the phase relationships between $\Phi$ and $\phi$ for small and large values of $\mu$, respectively. We conclude that the near-optimal phase-matching conditions are consistent with Eqs. (\ref{eq17}) and (\ref{eq18}).

\begin{figure}[htbp]
\begin{center}
\centering
\includegraphics[width = 16 cm]{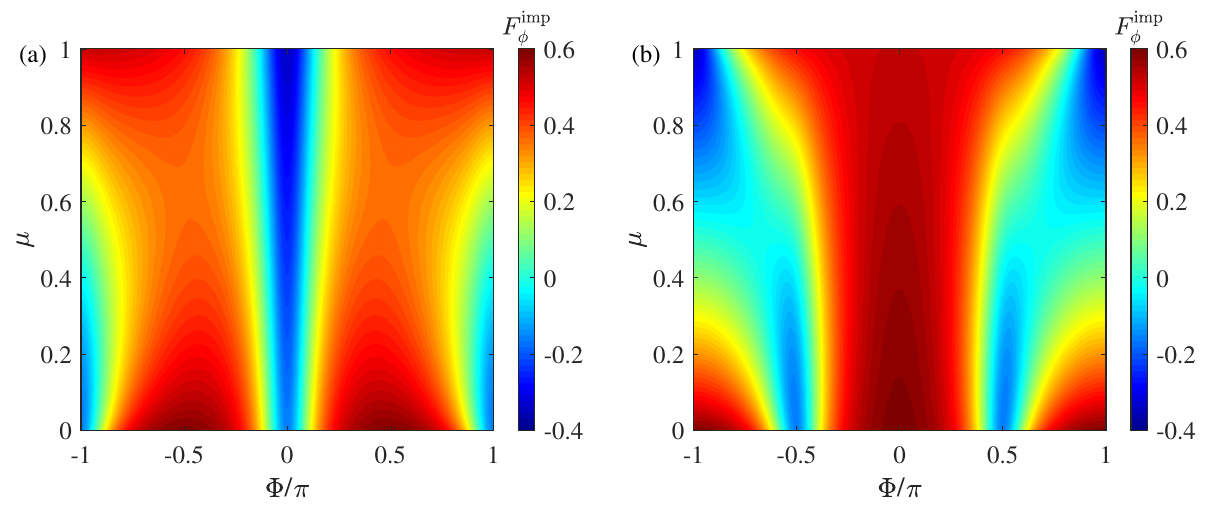}
\caption{(color online) $F_{\phi}^{\rm imp}$ as a function of $\Phi$ and $\mu$ with (a) $\phi=0$ and (b) $\phi=\pi/2$. The other parameters are $\gamma_{0}t=1$, $T=0.3$, $r=1$, and $\alpha=\frac{\sqrt{2}}{2}$.}
\label{fig5}
\end{center}
\end{figure}

Satisfying the phase-matching conditions outlined in Eqs. (\ref{eq17}) and (\ref{eq18}) will enhance the efficacy of utilizing squeezing to improve the estimation precision of the parameter $\phi$ within the squeezed reservoir. Failure to meet these conditions may degrade the phase sensitivity. This could be observed from Figs. \ref{fig4}(a) and \ref{fig4}(b) where the blue regions indicate negative values of $F_\phi^{\rm imp}$. It is essential to emphasize that, although we have derived only two representative phase-matching conditions for small and large values of $\mu$, as illustrated in Fig. \ref{fig4}, the general phase-matching condition is contingent upon the correlation factor and cannot be explicitly determined. An intuitive understanding is that mixed terms do not yield a definitive phase-matching condition, as the phase relationships are also mixed.
The numerical results presented in Fig. \ref{fig5} corroborate this extrapolation. It is evident from Fig. \ref{fig5} that the phase relationship becomes ambiguous when the mixed terms $\bar{\mu}\mu$ cannot be neglected; conversely, in the limiting cases, the phase-matching conditions are well-defined.

\section{Enhancing the Independent Estimation of Correlation Factor}
\label{mu}
Accurately estimating the correlation factor facilitates better utilization of channel correlations in the implementation of quantum information tasks.
In this section, we turn to consider how to enhance the precision of the correlation factor $\mu$. The QFI $F_\mu$ can be calculated as
\begin{eqnarray}
\label{eq20}
{{F}_{\mu }}=\frac{2\rho_{22}^{(u)}}{{\bar{\mu }}}+\frac{{{(\rho_{11}+\rho_{44})}^{2}}{({\rho_{22}^{(u)}})^{2}}+{{\left| {{\rho }_{\text{41}}} \right|}^{\text{2}}}\cdot 4\left[ \left(\rho_{11}\rho_{44}-|\rho_{41}|^{2}\right){{\left| {{D}_{\mu }} \right|}^{\text{2}}}+{{\left| {{\rho }_{\text{41}}} \right|}^{\text{2}}}{[{\operatorname{Re}(D_{\mu})}]^{2}}-(\rho_{11}+\rho_{44})\rho_{22}^{(u)} \operatorname{Re}(D_{\mu}) \right]}{(\rho_{11}+\rho_{44})\left(\rho_{11}\rho_{44}-{{\left| {{\rho }_{\text{41}}} \right|}^{\text{2}}}\right)},
\end{eqnarray}
where $D_{\mu}=(\rho_{41}^{(c)} - \rho_{41}^{(u)})/\rho_{41}$ physically represents the contrast between the relevant coherent terms and the irrelevant coherent terms.

\begin{figure}[htbp]
	\centering
\includegraphics[width = 16 cm]{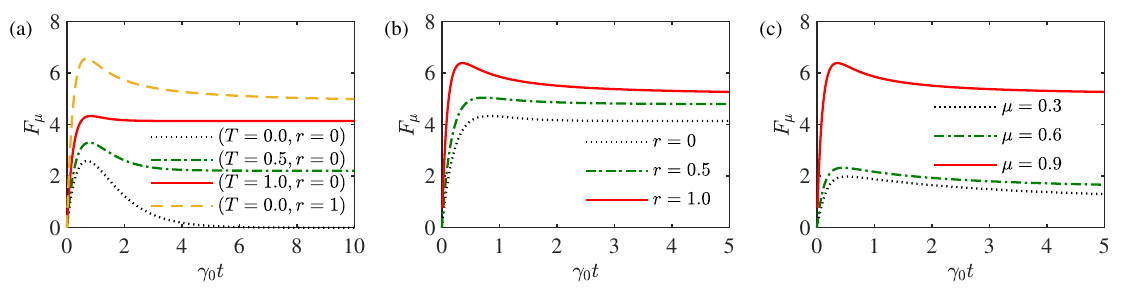}
 \caption{(color online) $F_{\mu}$ as a function of $\gamma_{0}t$ with (a) $\mu=0.9$, (b) $T={1}$, $\mu=0.9$, and (c) $T={1}$, $r=1$. The other parameters are $\Phi=\pi$, $\alpha=\frac{\sqrt{2}}{2}$, and $\phi=\pi/2$.}
	\label{fig6}
\end{figure}

The numerical results in Fig. {\ref{fig6}}(a) demonstrate the dynamic evolution of $F_\mu$ as a function of $\gamma_{0}t$ with various environmental temperatures. In the absence of squeezing, $F_\mu$ initially exhibits an increasing trend, followed by a subsequent decay over time. This behavior can be attributed to the fact that the quantum system does not initially encode information regarding the correlation factor $\mu$. As the probe state interacts with the environment, it acquires information about $\mu$, leading to an enhancement in the value of $F_\mu$. Notably, in contrast to the behavior observed for $F_\phi$, an increase in environmental temperature can actually enhance $F_\mu$. This enhancement occurs because a higher temperature implies a larger population of thermal photons participating in the system-reservoir exchange, thereby imprinting the correlation signature more effectively onto the probe state.
However, as the evolution time progresses, $F_\mu$ ultimately decays to a stable value or even tends to zero, which depends on the characteristics of the reservoir. When the environment is in a zero-temperature vacuum state (without squeezing), the system will decay to its ground state, resulting in a total loss of information about $\mu$. In contrast, if the environment is in a thermal state or a squeezed vacuum state, the absorption processes involving single photons (in the absence of squeezing) or two-photon processes (in the presence of squeezing) will drive the system toward a non-trivial steady state. Consequently, the quantity $F_\mu$ will approach a stable value.
Furthermore, the dependence of $F_\mu$ on $\gamma_{0}t$ for various squeezing strengths $r$ is also examined. As depicted in Fig. \ref{fig6}(b), the squeezing effect further amplifies $F_\mu$; specifically, greater squeezing strength is related to a more pronounced enhancement of $F_\mu$. To elucidate the influence of the correlation effect on $F_\mu$, we present in Fig. \ref{fig6}(c) the dynamics of $F_\mu$ as a function of $\gamma_{0}t$ for various values of $\mu$. As depicted in Fig. \ref{fig6}(c), it can be observed that with an increasing correlation factor, the precision is enhanced more significantly.

\begin{figure}[htbp]
	\centering
\includegraphics[width = 16 cm]{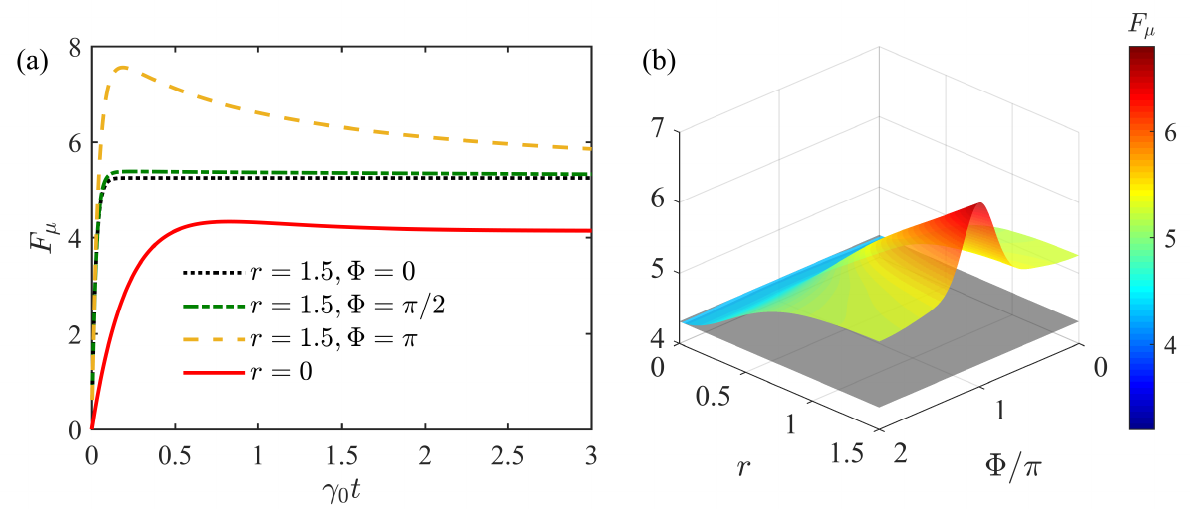}
\caption{(color online) The behaviors of $F_{\mu}$ in different cases. (a) $F_{\mu}$ as a function of $\gamma_{0}t$ for different squeezed phase $\Phi$. (b) The coloured surface presents $F_{\mu}$ as a function of $\Phi$ and $r$ at $\gamma_{0}t=1$, and the gray reference plane corresponds to the results obtained in a correlated thermal reservoir (without squeezing). The other parameters are $T=1$, $\mu={0.9}$, $\alpha=\frac{\sqrt{2}}{2}$, and $\phi=\pi/2$.}
	\label{fig7}
\end{figure}

Building on the discussions in Sec. \ref{phase}, we further investigate the influence of the squeezing phase $\Phi$ on the enhancement of $F_\mu$ via squeezing. As depicted in Fig. \ref{fig7}(a), $F_\mu$ is shown as a function of $\gamma_{0}t$ for various squeezing phases $\Phi$. The results indicate that a sufficiently large squeezing strength $r$ can enhance $F_\mu$ effectively for any value of $\Phi$, though the extent of this enhancement remains phase-sensitive.

A more comprehensive analysis is provided in Fig. \ref{fig7}(b), which presents $F_\mu$ as a function of both $r$ and $\Phi$ at a fixed time $\gamma_{0}t=1$. This confirms that, in comparison to a correlated thermal reservoir, the correlated squeezed thermal reservoir significantly enhances $F_\mu$ across the entire range of $r$ and $\Phi$. The observed variations in the degree of enhancement with respect to $\Phi$ underscore the necessity to identify the optimal phase-matching condition that maximizes $F_\mu$.

\begin{figure}[htbp]
	\centering
\includegraphics[width = 16 cm]{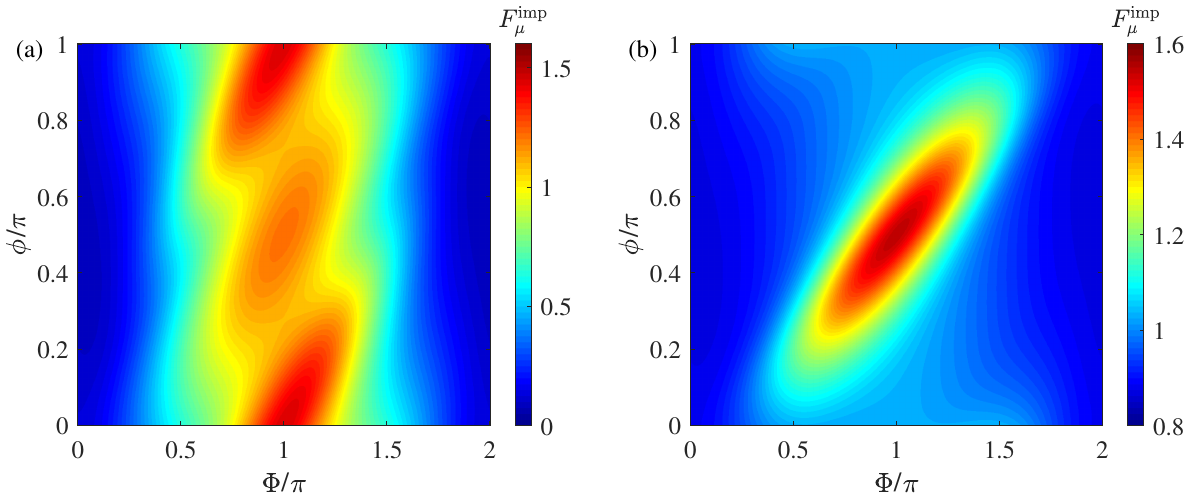}
\caption{(color online) $F_{\mu}^{\rm imp}$ as a function of $\Phi$ and $\phi$ with (a) $\mu=0.1$ and (b) $\mu=0.9$. The other parameters are $\gamma_{0}t=1$, $T=1$, $r=1$, and $\alpha=\frac{\sqrt{2}}{2}$.}
	\label{fig8}
\end{figure}

To determine the optimal squeezing phase, we can employ a method analogous to that described in Sec. \ref{phase}.
We define $F_\mu^{\rm imp}$ as follows
 \begin{equation}
  F_\mu^{\rm imp}=F_\mu-F_\mu(r=0).
  \label{eq21}
\end{equation}
Figure {\ref{fig8}}(a) illustrates the phase relationship between $\Phi$ and $\phi$ on $F_\mu^{\rm imp}$ for a small value of $\mu$. The corresponding phase-matching condition for $F_\mu^{\rm imp}$ is found to be
\begin{equation}
  \cos(2\Phi-2\phi)=1.
  \label{eq22}
\end{equation}
 Whereas for a large value of $\mu$, the corresponding phase-matching condition for $F_\mu^{\rm imp}$ can be obtained from Fig. {\ref{fig8}}(b), which reduces to
 \begin{equation}
 \cos(\Phi-2\phi)=1.
  \label{eq23}
 \end{equation}

\begin{figure}[htbp]
\begin{center}
\includegraphics[width=0.49\linewidth]{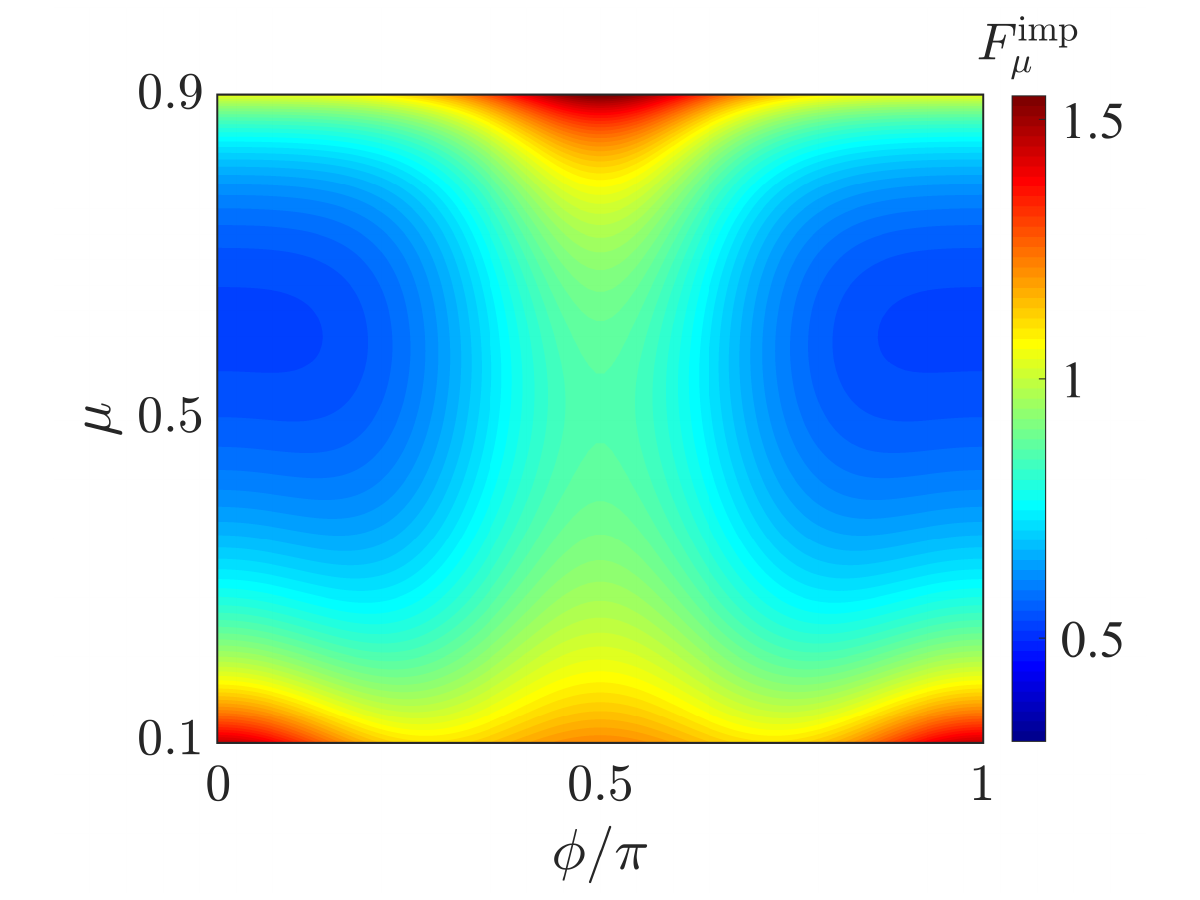}
\caption{(color online) $F_{\mu}^{\rm imp}$ as a function of $\phi$ and $\mu$. The other parameters are $\Phi=\pi$, $\gamma_{0}t=1$, $T=1$, $r=1$, and $\alpha=\frac{\sqrt{2}}{2}$.}
\label{fig9}
\end{center}
\end{figure}

It should be particularly noted that a moderate value of $\mu$ does not yield a definitive phase-matching condition. The behavior of $F_\mu^{\rm imp}$ as a function of $\mu$ and  $\phi$ is plotted in Fig. {\ref{fig9}}. One observation is that it verifies the correctness of Eqs. (\ref{eq22}) and (\ref{eq23}) in describing the phase-matching conditions for small and large values of $\mu$. Another feature of Fig. {\ref{fig9}} is that, even under the conditions of phase mismatch, $F_\mu^{\rm imp}$ remains positive, indicating an enhancement of $F_\mu$. Meeting the phase-matching conditions will be instrumental in maximizing the potential of squeezing to improve the estimation of the parameter $\mu$ within the context of a correlated squeezed reservoir.

\section{Enhancing the Joint Estimation of Phase and Correlation Factor}
\label{joint}
Quantum resources are both valuable and susceptible to decoherence, making it essential to minimize their consumption. A compelling strategy is to extract information about all parameters of interest within a single experiment. Joint estimation (also named as multiparameter estimation) protocols achieve this by reducing the resource requirements approximately by a factor equal to the number of parameters, compared to independent estimation schemes \cite{Yousefjani2017}. Notably, joint estimation introduces inherent trade-offs: optimizing the precision of one parameter often compromises the precision of others.

In joint estimation, a key objective is to achieve, for each parameter, a precision comparable to that attainable in an optimal independent estimation. This ideal scenario is realized under the condition of parameter compatibility \cite{Ragy2016}, which requires satisfying three criteria: (i) A single probe state $\varrho$ must suffice for all parameters, replacing the individual states ${\varrho_i}$ without loss of precision for any parameter; (ii) A single measurement strategy $\Pi_{\vec{x}}$ (where $\vec{x}$ denotes a vector of outcomes) must replace all individual measurements $\Pi_{x_i}$, while still guaranteeing optimal precision for each parameter; (iii) For the estimates to be mutually independent, the off-diagonal elements of the covariance matrix must vanish, ensuring that estimation errors for different parameters do not interfere. When these compatibility conditions are met, the joint estimation protocol achieves maximal efficiency --- reducing resource consumption without sacrificing estimation precision for any parameter.

In this section, we concentrate on the joint estimation of the correlation factor and the phase parameter. According to Eq. (\ref{eq12}), the off-diagonal element of the QFIM ($F_{\mu\phi}$), which quantifies the correlations between the estimators for $\phi$ and $\mu$ can be calculated as
\begin{eqnarray}
  {{F}_{\mu\phi}} &=&\frac{{{\left| {{\rho }_{\text{41}}} \right|}^{\text{2}}}\cdot \left[4\left(\rho_{11}\rho_{44}-|\rho_{41}|^{2}\right)\operatorname{Re}({{D}_{\phi }}D_{\mu }^{*})+\text{4}{{\left| {{\rho }_{\text{41}}} \right|}^{\text{2}}}\operatorname{Re}(D_{\phi})\operatorname{Re}(D_{\mu})-2(\rho_{11}+\rho_{44})\rho_{22}^{(u)} \operatorname{Re}(D_{\phi}) \right]}{(\rho_{11}+\rho_{44})\left(\rho_{11}\rho_{44}-{{\left| {{\rho }_{\text{41}}} \right|}^{\text{2}}}\right)},
\end{eqnarray}
where the cross-term $\operatorname{Re}(D_{\phi}D_{\mu}^{*})$ represents the interference between the phase and correlation sensitivities.
Note that $F_{\mu\phi}$ shares the same denominator structure as the individual QFIs in Eqs.~(\ref{eq13}) and (\ref{eq20}), revealing the intricate coupling between the two parameters governed by the common coherence factor $|\rho_{41}|^2$.

In the independent estimation scenario, the lower bound on the total variance of estimating the parameters $\phi$ and $\mu$ is given by the sum of the inverses of the corresponding diagonal elements of the QFIM,
\begin{equation}\label{eq25}
  \Delta_{\rm ind}=\sum\limits_{i}{F_{ii}^{-1}}=\frac{1}{F_{\mu}}+\frac{1}{F_{\phi}}.
\end{equation}

In contrast, in the joint estimation scenario, the lower bound on the total variance is determined by the trace of the inverse of the QFIM,
\begin{equation}
  \Delta_{\rm sim}=\frac{1}{2}\text{Tr}({F^{-1}})=\frac{1}{2}\cdot\frac{F_{\mu}+F_{\phi}}{F_{\mu}F_{\phi}-F_{\mu\phi}^{2}}.
  \label{eq26}
\end{equation}

To directly compare the estimation performance, we evaluate the ratio of the minimal total variances achieved under independent and joint estimation schemes \cite{Yousefjani2017}
\begin{equation}\label{eq27}
 R=\frac{\Delta_{\rm ind}}{\Delta_{\rm sim}}=2\cdot\frac{F_{\mu}F_{\phi}-F_{\mu\phi}^{2}}{F_{\mu}F_{\phi}},
\end{equation}
where $R \leq 2$ when only two parameters are considered. $R > 1$ signifies that there is a significant advantage of joint estimation for $\phi$ and $\mu$ over independent estimation strategies.

\begin{figure}[htbp]
	\centering
\includegraphics[width = 16 cm]{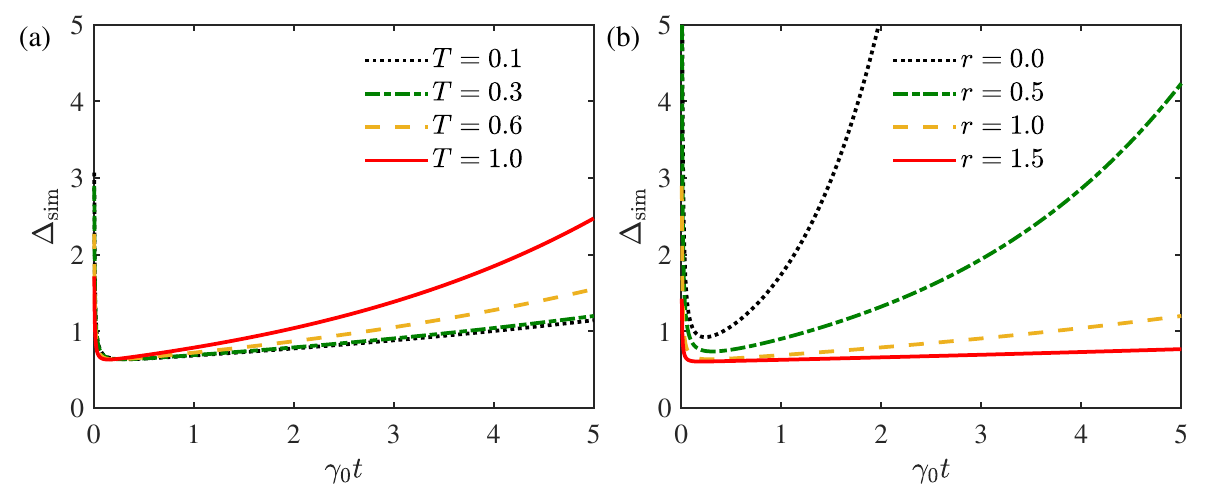}
\caption{(color online) $\Delta_{\rm sim}$ as a function of $\gamma_{0}t$ with (a) $r=1.0$ and (b) $T=0.3$. The other parameters are $\mu=0.9$, $\Phi=0$, $\alpha=\frac{\sqrt{2}}{2}$, and $\phi=\pi/2$.}
	\label{fig10}
\end{figure}

Figure {\ref{fig10}}(a) shows $\Delta_{\rm sim}$ as a function of $\gamma_0 t$ for different temperatures. The numerical results indicate that higher temperatures increase the variance of the joint estimation of parameters $\phi$ and $\mu$, thereby reducing its precision. Although elevated temperatures accelerate the decay of $F_{\phi}$, they also improve the precision of estimating $\mu$ via enhanced
$F_{\mu}$. Nevertheless, our analysis reveals that the temperature dependence of $\Delta_{\rm sim}$ is dominated by $F_{\phi}$, implying that the joint estimation precision is primarily limited by the sensitivity of the $\phi$ parameter. This occurs because
$F_{\mu}$ consistently attains larger values. As derived in Eq. (\ref{eq26}), the lower bound of
$\Delta_{\rm sim}$ is chiefly determined by $F_{\phi}$.

As illustrated in Fig. {\ref{fig10}}(b), squeezing reduces the variance and improves the joint estimation precision of
$\phi$ and $\mu$. Moreover, larger squeezing strength leads to a more substantial reduction in $\Delta_{\rm sim}$. It is also apparent that the squeezing phase plays a non-negligible role in the squeezing-enhanced estimation precision. Therefore, identifying the optimal squeezing phase-matching condition for minimizing $\Delta_{\rm sim}$ is of significant interest.

\begin{figure}[htbp]
	\centering
\includegraphics[width = 16 cm]{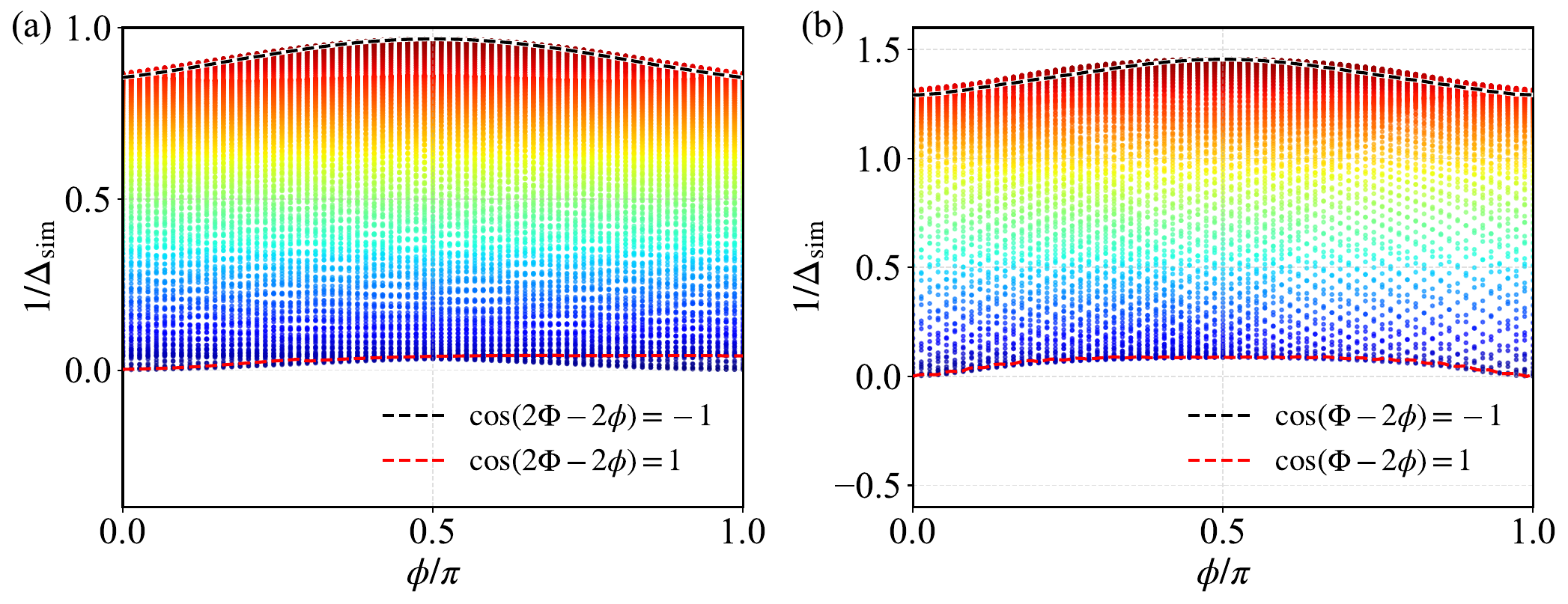}
\caption{(color online) $1/\Delta_{\rm sim}$ as a function of $\phi$ with the squeezing phase $\Phi$ varying from $0$ to $2\pi$  at (a) $\mu=0.1$ and (b) $\mu=0.9$. The black and red curves denote $1/{\Delta_{\rm sim}}$ under the near-optimal phase-matching conditions with respect to $\phi$ and $\mu$, respectively. The other parameters are $\gamma_{0}t=1$, $T=0.3$, $r=1$, and $\alpha=\frac{\sqrt{2}}{2}$.}
	\label{fig11}
\end{figure}

Given that the analytical expression for $\Delta_{\rm sim}$ is complex and offers limited intuitive insight, we turn to numerical results to clarify the phase relationship between $\Phi$ and $\phi$. Figure {\ref{fig11}} illustrates the joint precision $1/\Delta_{\rm sim}$ as a function of $\phi$ with the squeezing phase varying from 0 to $2\pi$. The black and red curves denote $1/{\Delta_{\rm sim}}$ under the near-optimal phase-matching conditions with respect to $\phi$ and $\mu$, respectively.
Evidently, in the weak correlation regime (small $\mu$), the corresponding near-optimal squeezing phase-matching condition for $1/\Delta_{\rm sim}$ can be inferred from Fig. {\ref{fig11}}(a),
\begin{equation}
  \cos(2\Phi-2\phi)=-1.
  \label{eq28}
\end{equation}

Conversely, for strong correlations (large $\mu$), the corresponding phase-matching condition for $1/\Delta_{\rm sim}$ can be obtained from Fig. {\ref{fig11}}(b),
 \begin{equation}
  \cos(\Phi-2\phi)=-1.
  \label{eq29}
 \end{equation}

It is found that the near-optimal phase-matching condition for maximizing $1/\Delta_{\rm sim}$ is equivalent to that for maximizing $F_\phi$. As shown in Fig. {\ref{fig11}}, the optimal phase-matching condition representing $\phi$ (black curves) essentially coincides with the maximum value of $1/\Delta_{\rm sim}$, while the optimal phase-matching condition representing $\mu$ (red curves) approximately aligns with the minimum value of $1/\Delta_{\rm sim}$. This behavior stems from the fact that the individual optimal phase-matching conditions for estimating parameters $\phi$ and $\mu$, as derived in Sec. \ref{phase} and Sec. \ref{mu}, do not coincide with each other. Within the joint estimation scheme, we therefore adopt a trade-off strategy that prioritizes optimization with respect to $\phi$. This choice is motivated by the observation that optimizing for $\mu$ accelerates the decay of $F_\phi$, leading to a sharp increase in the joint estimation variance. In contrast, when the protocol is optimized primarily for $\phi$, $F_\mu$ remains consistently large across all squeezing phases, thereby helping to maintain a low joint estimation variance despite the inherent trade-off. Hence, the phase-matching condition that minimizes the joint variance is equivalent to that which enhances $F_\phi$.

\begin{figure}[htbp]
	\centering
\includegraphics[width = 16 cm]{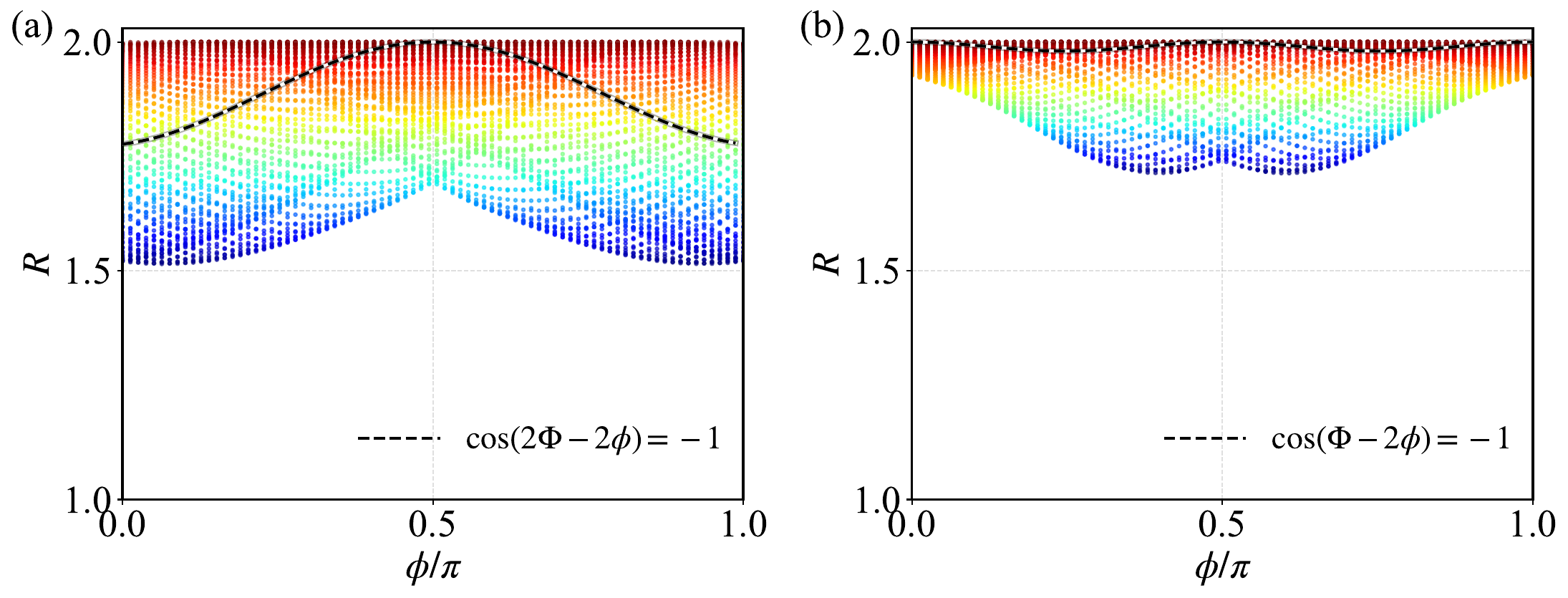}
 \caption{(color online) $R$ as a function of $\phi$ with the squeezing phase $\Phi$ varying from $0$ to $2\pi$ at (a) $\mu=0.1$ and (b) $\mu=0.9$; The black curves denote $R$ under the near-optimal phase-matching conditions with respect to $\phi$. The other parameters are $\gamma_{0}t=1$, $T=0.3$, $r=1$, and $\alpha=\frac{\sqrt{2}}{2}$.}
	\label{fig12}
\end{figure}

Although the parameters $\phi$ and $\mu$ are incompatible - failing to satisfy conditions (ii) and (iii) - simultaneous estimation of both parameters can still yield a net advantage. Figure {\ref{fig12}} displays the performance metric $R$ as a function of $\phi$ with the squeezing phase varying from 0 to $2\pi$, where $R \geq 1$ indicates the superiority of joint over independent estimation. Remarkably, this advantage persists across all values of $\Phi$ and $\phi$. One might anticipate that $R$ reaches its maximum when the phase-matching conditions (\ref{eq28}) and (\ref{eq29}) are met. However, it is important to emphasize that a maximum in $R$ does not imply minimization of the joint estimation variance $\Delta_{\text{sim}}$, since $R$ also depends on the variance $\Delta_{\text{ind}}$ from independent estimations. This explains why the maximum value of $R$ does not necessarily occur under the phase-matching conditions, as evidenced in Fig. {\ref{fig12}}.

\section{Conclusions}
 \label{conclusions}

In conclusion, we have investigated the potential of correlated squeezed-thermal reservoirs for enhancing quantum parameter estimation, with particular emphasis on the role of the squeezing phase---a factor frequently overlooked in previous studies---in the estimation of the quantum phase parameter and the correlation factor. We find that the choice of squeezing phase is crucial for achieving high precision in both individual and joint estimation protocols. Furthermore, we demonstrate that in reservoirs exhibiting correlation effects, the phase-matching condition required for squeezing to enhance estimation precision depends on the correlation strength $\mu$. For clarity, we summarize these near-optimal phase-matching conditions for maximizing $F_\phi$, $F_\mu$, and minimizing $\Delta_{\text{sim}}$, which form the central results of our work, in Table \ref{table1}.

\begin{table}[htbp]
  \centering
  \caption{The near-optimal phase-matching conditions for maximizing $F_\phi$, $F_\mu$, and minimizing $\Delta_{\text{sim}}$, where $\Phi$ denotes the squeezing phase, $\phi$ represents the unknown phase parameter as defined in Eq. (\ref{eq7}) and $\mu\in[0,1]$ is the correlation parameter.}
  \label{table1}
  \begin{tabular}{|c|c|c|}
    \hline
    & small $\mu$ & large $\mu$ \\ \hline
    independent $\phi$ & $\cos(2\Phi-2\phi)=-1$ & $\cos(\Phi-2\phi)=-1$ \\ \hline
    independent $\mu$  & $\cos(2\Phi-2\phi)=1$ & $\cos(\Phi-2\phi)=1$ \\ \hline
    joint $\phi$ and $\mu$ & $\cos(2\Phi-2\phi)=-1$ & $\cos(\Phi-2\phi)=-1$ \\ \hline
  \end{tabular}
\end{table}

Notably, the condition for minimizing the joint estimation variance coincides with that which enhances the precision of the phase parameter $\phi$. This is due to the fact that squeezing significantly degrades the precision of $\phi$ under phase mismatch, leading to a sharp increase in the total variance of simultaneous estimation. In contrast, the precision of estimating the correlation factor $\mu$ benefits from squeezing even in the absence of phase matching. We also compare the advantages of joint and independent estimation strategies. While independent estimation can achieve higher precision for each parameter individually, joint estimation improves the efficiency of quantum resource utilization while maintaining high overall precision. This work enhances the understanding of key prerequisites for exploiting squeezing and correlation effects to optimize parameter estimation in correlated squeezed reservoirs, offering valuable insights for the engineering of such reservoirs in quantum information technologies.

\section*{CRediT authorship contribution statement}
\textbf{Cai-Hong Liao:} Calculating, Data curation, Writing.
\textbf{Yan-Ling Li:} Conceptualization, Methodology, Data curation, Writing original draft.
\textbf{Long Huang:} Calculating, Data curation.
\textbf{Xing Xiao:} Methodology, Writing and Editing, Supervision.

\section*{Declaration of competing interest}
The authors declare that they have no known competing financial interests or personal relationships that could have appeared to influence the work reported in this paper.

\section*{Data availability}
No data was used for the research described in the article.

\section*{Acknowledgements}
We would like to express our gratitude to Prof. Qing-Shou Tan for his helpful discussions. X. Xiao is supported by the National Natural Science Foundation of China (Grant Nos. 12265004, 12534020) and the Natural Science Foundation of Jiangxi Province (Grant No. 20242BAB26010). Y.L. Li is supported by the National Natural Science Foundation of China (Grant No. 12365003) and Jiangxi Provincial Key Laboratory of Multidimensional Intelligent Perception and Control of china (No. 2024SSY03161).

%\appendix
%\section{Example Appendix Section}
%\label{app1}

%%       \citet{<label>} ==> Lamport (1994)
%%       \citep{<label>} ==> (Lamport, 1994)
%%
%Example citation, See \citet{lamport94}.



%\section*{References}

\end{document}